\documentclass[sigconf]{acmart}

\usepackage{booktabs, balance}
\usepackage{dcolumn,multirow}
\usepackage{textcomp}
\usepackage[utf8]{inputenc}
\usepackage[T1]{fontenc}
\usepackage[T1]{fontenc}
\usepackage{hyperref}
\usepackage{todonotes} 
\usepackage{array}
\usepackage{listings}
\usepackage{tcolorbox}
\usepackage{enumitem}
\usepackage{multirow}
\usepackage{makecell}
\usepackage{natbib}
\usepackage{tikz}
\usepackage{pgfplots}
\usepackage{pgfplotstable}
\usepackage{longtable}
\usetikzlibrary{positioning}
\usetikzlibrary{arrows,decorations.pathreplacing,calligraphy}
\newcommand{\tablehack}{\vspace{-12pt}}

\usepackage{multicol}

\usepackage{float}

\newcommand{\itemizehack}{\vspace{-4pt}}
\newcommand{\subsectionhack}{\vspace{-4pt}}
\newcommand{\figendhack}{\vspace{-6pt}}
\newcommand{\ignore}[1]{}

\AtBeginDocument{%
  \providecommand\BibTeX{{%
    \normalfont B\kern-0.5em{\scshape i\kern-0.25em b}\kern-0.8em\TeX}}}

\setcopyright{acmcopyright}
\copyrightyear{2018}
\acmYear{2018}
\acmDOI{XXXXXXX.XXXXXXX}

\acmConference[Conference acronym 'XX]{Make sure to enter the correct
  conference title from your rights confirmation emai}{June 03--05,
  2018}{Woodstock, NY}
  
\acmBooktitle{Woodstock '18: ACM Symposium on Neural Gaze Detection,
 June 03--05, 2018, Woodstock, NY} 
\acmISBN{978-1-4503-XXXX-X/18/06}

\begin{document}

\title{CS1-LLM: Integrating LLMs into CS1 Instruction}


\author[Annapurna]{Annapurna Vadaparty}
\affiliation{
  \institution{University of California, San Diego}
  \country{USA}
}
\email{avadaparty@ucsd.edu}

\author[Daniel]{Daniel Zingaro}
\affiliation{
  \institution{University of Toronto Mississauga}
  \country{Canada}
}
\email{daniel.zingaro@utoronto.ca}

\author[David]{David H. Smith IV}
\affiliation{
  \institution{University of Illinois}
  \city{Urbana}
  \country{USA}
}
\email{dhsmith2@illinois.edu}

\author[Mounika]{Mounika Padala}
\affiliation{
  \institution{University of California, San Diego}
  \country{USA}
}
\email{mpadala@ucsd.edu}

\author[Christine]{Christine Alvarado}
\affiliation{
  \institution{University of California, San Diego}
  \country{USA}
}
\email{cjalvarado@ucsd.edu}

\author[Jamie]{Jamie Gorson Benario}
\affiliation{
  \institution{Google}
  \country{USA}
}
\email{jamben@google.com}

\author[Leo]{Leo Porter}
\affiliation{
  \institution{University of California, San Diego}
  \country{USA}
}
\email{leporter@ucsd.edu}

\ignore{ 
\author{Ben Trovato}
\email{trovato@corporation.com}
\orcid{1234-5678-9012}
\author{G.K.M. Tobin}
\authornotemark[1]
\email{webmaster@marysville-ohio.com}
\affiliation{%
  \institution{Institute for Clarity in Documentation}
  \streetaddress{P.O. Box 1212}
  \city{Dublin}
  \state{Ohio}
  \country{USA}
  \postcode{43017-6221}
}
} 

\renewcommand{\shortauthors}{Vadaparty et al.}

\begin{abstract}
The recent, widespread availability of Large Language Models (LLMs) like ChatGPT and GitHub Copilot may impact introductory programming courses (CS1) both in terms of what should be taught and how to teach it.  Indeed, recent research has shown that LLMs are capable of solving the majority of the assignments and exams we previously used in CS1.  In addition, professional software engineers are often using these tools, raising the question of whether we should be training our students in their use as well.  This experience report describes a CS1 course at a large research-intensive university that fully embraces the use of LLMs from the beginning of the course.  To incorporate the LLMs, the course was intentionally altered to reduce emphasis on syntax and writing code from scratch. Instead, the course now emphasizes skills needed to successfully produce software with an LLM. This includes explaining code, testing code, and decomposing large problems into small functions that are solvable by an LLM. In addition to frequent, formative assessments of these skills, students were given three large, open-ended projects in three separate domains (data science, image processing, and game design) that allowed them to showcase their creativity in topics of their choosing.  In an end-of-term survey, students reported that they appreciated learning with the assistance of the LLM and that they interacted with the LLM in a variety of ways when writing code. We provide lessons learned for instructors who may wish to incorporate LLMs into their course.

\end{abstract}

\ccsdesc[500]{Social and professional topics~Computer science education}

\keywords{CS1, Introductory Programming, LLM, Copilot, Generative AI}



\maketitle

\section{Introduction}

In 2023, the computing education research community witnessed an explosion of commentary and research on the impact of Large Language Models (LLMs) and Generative AI (GenAI) on computing education courses. In an ITiCSE Working Group report published in December 2023, the authors note ``There is little doubt that LLMs and other forms of GenAI will have a profound impact on computing education over the coming years''~\cite{iticse23wg}. Indeed, this profound impact is already being felt. For example, much research has demonstrated that GPT-4 can solve CS1 problems at the level of a top student in the course~\cite{iticse23wg}. Capabilities of LLMs are increasing rapidly, leading instructors to worry about what their programming courses should look like now. Some instructors are attempting to ban the use of the tools, or devising types of assessments where LLMs struggle, while others are embracing the changes ~\cite{guo23}.

We suggest a redesign of introductory programming courses around LLMs, including reprioritizing learning outcomes, for two reasons.
First, with the increasing capability of LLMs, the practice of software development and the required skills for programming are evolving. In a survey from GitHub, it was reported that 92\% of developers in the US are using these tools and that 70\% of those developers are seeing benefits such as upskilling and increased productivity~\cite{github_survey}. We argue that students should be learning with LLMs to help them prepare for a work force that is using LLMs and thus skills we once prioritized, like writing code from scratch, may no longer be as relevant. 
The second reason is that incorporating LLMs can allow students to more quickly engage in larger, open-ended projects which have more personal relevance to them and can improve engagement~\cite{hulls2015use}. CS1 students typically work on small, highly constrained problems that do not resemble the software development process ~\cite{allen2019analysis}. By working on open-ended projects, they gain better exposure to programming practices and are motivated by working on what may be more personally relevant~\cite{sharmin2020weekly}.

In this paper, we discuss our experiences teaching a new CS1 (CS1-LLM) with LLMs at the center of learning to program. Our core question in designing the course was: What can students do now with LLMs that they could not do before? We still want students to be able to write code from scratch, and we remain committed to other fundamentals~\cite{venables2009closer} such as carefully reading, tracing, and explaining code. But rather than spending several weeks teaching Python syntax as we have done in the past, we now rely on the LLM's powerful code-writing abilities to help students overcome syntax hurdles and focus on more creative aspects of programming. 

\subsectionhack
\subsectionhack
\section{Design Principles}

As mentioned in the previous section, we designed our new CS1 (CS1-LLM) to help students benefit from the affordances of LLMs. With that overarching goal in mind, we began by establishing the design principles that would guide the development and implementation of our course.

Our first design principle is to enable and encourage students to use LLMs throughout their coursework. To that end, we taught students early how to install and use GitHub Copilot to help them write Python code. We chose Copilot as it integrates well with VS Code, an IDE that many of our students will use in industry or their later programming projects. As we wanted to test students in a setting mirroring the setting in which they learned, we additionally made Copilot available on some supervised quizzes and tests. There is lack of consensus on what the community believes constitute fundamentals that should be learned with and without LLMs. We therefore aimed to ensure that students would emerge from the course with fluency in coding both with and without LLM support. A key skill that is expected after completion of CS1 is to be able to independently read, trace, and write some code, so on some assessments we did not allow the use of Copilot. 

Our second design principle is to prepare students to enter a workforce where the use of LLMs will be the norm. It is unfortunately the case that there is a large gap between what is taught in CS courses and the needs of industry~\cite{craig2018listening,valstar2020quantitative}. For example, researchers have found that assignments in school tend to ask students to write code from scratch rather than having students learn to read and modify code as is commonly done in industry work, and that school assignments often focus on writing small standalone programs rather than adding features to existing programs~\cite{craig2018listening}. To reduce this gap, we not only teach students to use LLMs, but also use the affordances of LLMs to enable students to develop code reading and modification skills.

Our third design principle is to use what we know from research to improve outcomes for students from underrepresented groups. While not specific to LLMs, we employed several best practices known to improve student outcomes particularly for students from underrepresented groups~\cite{salguero2020longitudinal}. These practices include Peer Instruction (PI)~\cite{crouch2001peer,porter2016multi}, media computation~\cite{guzdial2003media}, and pair programming~\cite{mcdowell2006pair}. We also wanted to ensure students had the ability to receive frequent formative feedback, both through Peer Instruction~\cite{porter2016multi} and practice problems~\cite{west2015prairielearn}.

Our fourth design principle is to provide opportunities for students to be creative. There is a tendency in CS1 courses to use small, highly constrained problems~\cite{allen2019analysis} that can be auto-tested using predefined test cases. Unfortunately, those assignments do not provide opportunities for students to work on the types of open-ended projects that inspire creativity and better resemble industry~\cite{sharmin2021creativity}. We have changed our assignments to enable student choice in their programming through open-ended projects.

Our fifth and final design principle is to build a course that will equally serve both students who continue in CS and those who will not take any further computing courses.  We want our students to be able to both build on the computing principles they learn and apply those principles to their work even outside of the computing discipline. This necessarily means we need to spend less time on low-level syntax to make room for students to work at higher levels where meaningful work can be done. For example, we introduce and encourage students to use powerful Python modules for automating tedious tasks (such as merging a huge number of pdf files or identifying duplicate images in a huge image library).

\subsectionhack
\subsectionhack
\section{Course Context}

The CS1 course took place in a research-intensive university in North America.  The course included 10 weeks of instruction and one week for final exams. In the curriculum at the university, this ``CS1'' course is designed for students with no prior programming experience and is really the first half of two 10-week courses that are equivalent to a 10-week accelerated CS1 course for students with prior programming experience.  The course has 3 hours of weekly in-person, instructor-led classes; 1 hour per week of in-person discussion (led by a graduate Teaching Assistant (TA)); and 1 hour per week of in-person closed labs (also led by a TA) where students complete a programming activity, often in pairs.  The instructional staff consists of 5 graduate TAs and 33 undergraduate Instructional Assistants.  The course redesign team included graduate students and faculty from two institutions as well as a member of the software engineering industry.  Use of student data for this work is approved by our Human Subjects Board.

The course has a diverse student population with students from many different majors of study enrolled in the course.  The demographics of the course can be found in Table~\ref{tab:course_demographics}. All data except number of Computing Majors came from the end-of-term survey completed by 315 out of 552 students (57\% participation rate); Computing Majors came from the course pre-survey completed by 80\% of the students enrolled at the start.  ``Computing Majors'' includes Computer Science, Computer Engineering, Math with Emphasis in Computer Science, Bioinformatics, and Data Science.  Note that in the United States, Pell-Grant Eligibility can be seen as a proxy for low-income students as this denotes eligibility for federal financial aid.


\begin{table}[!ht]
\centering
\caption{Course Demographics\tablehack}
\label{tab:course_demographics}
\begin{tabular}{@{}p{4cm}ccc@{}}
\toprule
\multicolumn{1}{c}{\textbf{\begin{tabular}[c]{@{}c@{}}Group\end{tabular}}} & \textbf{\begin{tabular}[c]{@{}c@{}}Yes \end{tabular}} & \textbf{\begin{tabular}[c]{@{}c@{}}No \end{tabular}} & \textbf{Decline } \\ \midrule
Computing Major  & 33.7\%       & 66.3\%      & --  \\ 
\vspace{-6pt}
\textbf{Gender} \\
Male &  44.3\% & 52.2\% & 3.5\% \\
Female   & 50\%  & 46.5\%  & 3.5\%    \\
Nonbinary & 2.2\%   & 94.3\%   & 3.5\%  \\
\vspace{-6pt}
\textbf{Race/Ethnicity}\\
Hispanic or Latine & 27.0\% & 70.2\% & 2.9\% \\ 
Native American & 2.5\%    & 77.8\% & 19.7\%  \\
Black or African American & 3.2\% & 77.1\%  & 19.7\%  \\
East or Southeast Asian  & 43.8\%  & 36.5\%  & 19.7\%  \\
Indian or other South Asian  & 9.5\%   & 70.8\%  & 19.7\%  \\ 
Pacific Islander & 0.6\% & 79.7\% & 19.7\%  \\
North African/Middle-Eastern & 3.2\% & 77.1\% & 19.7\%  \\ 
White or Caucasian & 22.5\% & 57.8\% & 19.7\% \\ 
\vspace{-6pt}
\textbf{Student Status}\\
Transfer Student & 7.3\% & 91.1\% & 1.6\%  \\ 
First-Gen. College Student & 43.2\% & 50.8\% & 6.0\%  \\
Pell Grant Eligible & 47.0\% & 30.8\% & 22.2\% \\
\bottomrule
\end{tabular}
\end{table}
The course teaches variables, functions, conditionals, loops, \allowbreak strings, lists, and dictionaries in Python. The second half of the 2-part CS1 course, not otherwise discussed here, teaches classes, objects, inheritance, and poly\-morphism in Java.  Both before and after the transition to CS1-LLM, our course is taught using Peer Instruction and Live Coding in class and Pair Programming in labs, with at least two weeks spent on Media Computation.  The change to incorporate LLMs occurred for the Fall 2023 offering of the course.  The course was taught by an experienced instructor who has twelve years of experience and multiple teaching awards.

\section{Course Learning Goals}

Using our design principles, we revised the learning goals from the course and divided them by Bloom's taxonomy~\cite{krathwohl2002revision}. Key to our updates is the inclusion of learning goals specific to using LLMs.

\noindent \textbf{Level 1: Knowledge}
\begin{itemize}
    \itemizehack
    \item Define nondeterminism, Large Language Model (LLM), prompt, prompt engineering, code correctness, problem decomposition, and top-down design.
\end{itemize}

\noindent \textbf{Level 2: Comprehension}
\begin{itemize}
\itemizehack
    \item Illustrate the workflow that is used when programming with an AI assistant.
    \item Describe the purpose of common Python programming features, including variables, conditionals, loops, functions, lists, dictionaries, and modules.
\end{itemize}

\noindent \textbf{Level 3: Application}
\begin{itemize}
\itemizehack
    \item Apply prompt engineering to influence code generated by an AI assistant.
\end{itemize}

\noindent \textbf{Level 4: Analysis}
\begin{itemize}
\itemizehack
    \item Analyze and trace a Python program to determine or explain its behavior.
    \item Divide a programming problem into subproblems as part of top-down design.
    \item Debug a Python program to locate bugs.
\end{itemize}

\noindent \textbf{Level 5: Synthesis}
\begin{itemize}
\itemizehack
    \item Design open- and closed-box tests to determine whether code is correct.
    \item Identify and fix bugs in Python code.
    \item Modify Python code to perform a different task.
    \item Write complete and correct Python programs using top-down design, prompting, testing, and debugging.
\end{itemize}

\noindent \textbf{Level 6: Evaluation}
\begin{itemize}
\itemizehack
    \item Judge whether a program is correct using evidence from testing and debugging.
\end{itemize}

Some learning goals remain similar to a traditional CS1 course, in that students are still asked to read, trace, and explain code~\cite{venables2009closer}; debug code; modify code; and recognize common Python programming constructs.  Our CS1-LLM has less emphasis on writing code from scratch and fixing syntax as compared to our traditional CS1 class.  New learning goals for the CS1-LLM course include problem decomposition, top-down design, prompt engineering, and a larger emphasis on testing and debugging.

\subsectionhack
\subsectionhack
\section{Our Course--CS1-LLM}
\label{sec:our_course}

In redesigning the course we used a backwards design where we first consider the course learning goals and work backwards to design assessments and instruction that support students achieving those outcomes~\cite{wiggins2005understanding}.  This section describes the main components of the course that were adopted as part of reorienting the course instruction around LLMs. Ultimately, nearly every element of the course including lectures, assessments, and labs were changed for this new version of the course.  We have made course materials public in the hopes that others will be able to adopt and benefit from our work~\cite{CS1-LLM-materials}.

\subsection{Course Schedule}
\subsectionhack

\begin{figure}[ht]
    \centering
    \includegraphics[width=\columnwidth]{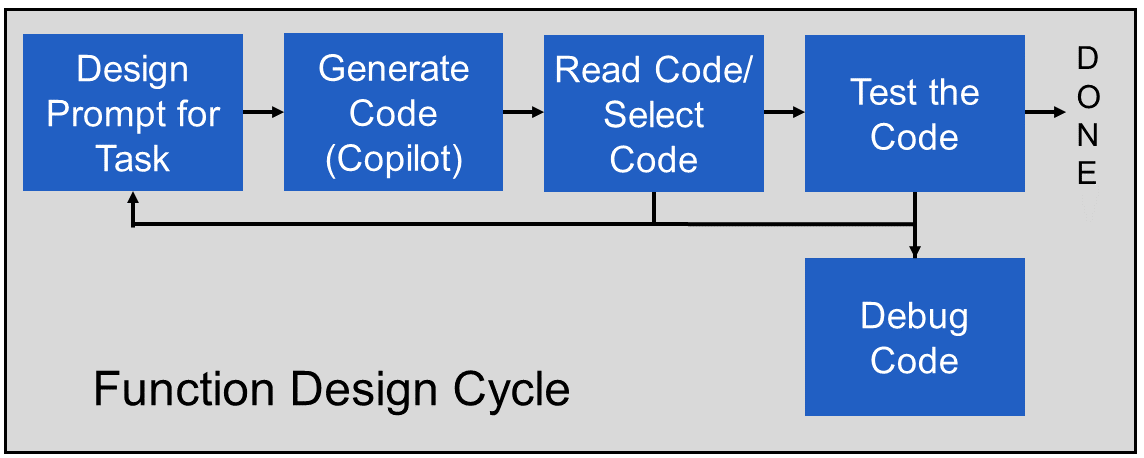}
    \caption{Workflow recommended to students for working with Copilot.  Based on workflow in course textbook~\cite{porter2024learn}.\figendhack}
    \label{fig:LLM_workflow}
\end{figure}

The topics by week in the course appear in Table~\ref{tab:course_schedule}. 
For the first four weeks, students focused on learning how to read, trace, and explain code; they also learned how to ask Copilot to explain code and to use the VSCode debugger to gain insight into the state of memory during program execution. Students learned the basics of variables, conditionals, loops, functions, strings, and lists.  The remaining 6 weeks taught the software engineering process when working with Copilot (see Figure~\ref{fig:LLM_workflow}), fleshing out ideas like testing, debugging, and problem decomposition in three separate domains.  For two weeks each, we taught students these concepts in the context of data science, image manipulation~\cite{guzdial2003media}, and games.  To ensure students had access to Copilot by the start of the project weeks, students had a mandatory assignment to set up VSCode with Copilot due in the second week of the class.  Students were also provided with videos about how to set up their system for both Windows and Mac.

\begin{table}[ht]
\caption{\tablehack Course Schedule}
\label{tab:course_schedule}
\begin{tabular}{|c|l|}
\hline
Week & Topic(s) \\
\hline \hline
1 & Functions and Working with Copilot\\ \hline
2 & Variables, Conditionals, Memory Models\\ \hline
3 & Loops, Strings, Testing, VSCode Debugger\\ \hline
4 & Loops, Lists, Files, Problem Decomposition \\ \hline
5 & Intro to Data Science, Dictionaries\\ \hline
6 & Revisit Problem Decomposition and Testing \\ \hline
7 & Intro to Images, PIL, Image Filters\\ \hline
8 & Copying Images, Intro to Games and Randomness\\ \hline
9 & Large Game Example\\ \hline
10 & Python Modules and Automating Tedious Tasks\\ \hline
\end{tabular}
\tablehack
\end{table}

\subsectionhack
\subsection{Textbook}
\subsectionhack

We used the recently published book ``Learn AI-Assisted Python Programming with GitHub Copilot and ChatGPT'' as the primary course textbook~\cite{porter2024learn}.  The book was selected because it teaches readers how to code with the aid of an LLM, had a games project chapter that overlapped well with our final project, and still teaches students how to understand Python code.

\subsectionhack
\subsection{Lectures}
\subsectionhack

Lectures consisted of a combination of mini-lectures, Live Coding, and Peer Instruction.  In addition to standard materials one might regularly see in a CS1 class, most lectures contained discussion of Copilot interactions. Students were given responses from Copilot that were incorrect for the task and were asked to identify the error with the response or write test cases that would uncover the mistake.  Lecture also showed conversations with Copilot Chat. Copilot Chat is similar to ChatGPT but it is integrated into VS Code and has access to the current code being written. These Copilot Chat conversations were used to help students dive deeper into understanding their code, or to offer examples of how students can use Chat to explore Python libraries that may help them complete a given task. When drawing memory diagrams for students, the instructor alternated between drawing on their tablet in class and using the VSCode Debugger to show the current state of execution.

\subsection{Assessments}
\subsectionhack

We designed a variety of new assessments that better align with the new learning goals and structure for the course.  

\noindent \textbf{Formative Assessments.}  Our third design principle sought to offer students frequent formative feedback.  In addition to in-class Peer Instruction questions and the course labs, we adopted PrairieLearn~\cite{west2015prairielearn} to offer students many opportunities to practice code tracing, code explaining, code testing, and code writing both on homework and as  practice quizzes.
Formative Assessments were worth 35\% and prepared students for the larger summative assessments. Peer Instruction participation was worth 5\%, reading quizzes 5\%, homework 15\%, and labs 10\%.  Summative assessments were worth 65\%, with projects worth 10\%, quizzes 30\%, and final exam 25\%.

\noindent \textbf{Homework.}  Students were given a homework on PrairieLearn~\cite{west2015prairielearn} each week.  The homework consisted of a variety of problem types (multiple choice, short answer, Explain in Plain English~\cite{smith2023code}, Parson's Problems~\cite{denny2008evaluating}, debugging, code writing) and students were allowed multiple attempts to solve each problem correctly.  The homework was designed to be completed without using an LLM but students were told they could use an LLM if stuck.

\noindent \textbf{Quizzes.}  The four 50-minute quizzes for the course increased in complexity as the term progressed. All four quizzes included code tracing, code explaining~\cite{venables2009closer}, Parsons Problems~\cite{denny2008evaluating}, and small code writing questions, mostly without access to Copilot.  In addition, on later quizzes, students were asked to answer questions on testing, debugging, and problem decomposition.
As the data science, image manipulation, and game domains were introduced in class, they were also introduced on quizzes.  

\noindent \textbf{Projects.}  Students completed one project per domain of data science, image manipulation, and games.  Each project was intentionally open-ended, allowing students to showcase their creativity.  The first project on data science asked students to find a dataset on Kaggle~\cite{kaggle}, ask a question that can be answered by that data, and then write a program to answer that question.  The second project on images asked students to write a program to create an image collage by filtering their images and pasting images on top (or adjacent) to one another.  The third project on games asked students to design a text-based game and implement either a playable game with user interaction or a game simulation to determine the likelihood of winning a particular game.  Students were allowed to use Copilot for the projects to increase the scale of what they could accomplish in a CS1 course. To help students properly scope their projects, they were required to meet with an instructional assistant a week before the due date to obtain guidance on proper scoping and get hints on how they might go about solving the problem.

We had students submit their code and any supporting files, a diagram of the functions they created when they decomposed the problem into multiple subproblems, and a 5-minute video of them explaining their project with at least 3 minutes explaining how one of their functions worked. Each project was graded by an instructional assistant and grading took 10--15 minutes per project.

\noindent \textbf{Labs.}  Each week students were given a lab to complete.  These labs were completed either during the synchronous mandatory 50-minute lab session or, on quiz days when the lab was used for a quiz, at home.  Labs were designed to help students get started with the programming concepts or programming domain (e.g., images) discussed in lecture that week.  Labs included working with Copilot to solve a problem, writing code without Copilot, and debugging buggy code using the VSCode Debugger.

\noindent \textbf{Final Exam.}  The final exam consisted of three parts: 1) (90 minutes, worth 70\% of the exam grade) a primarily multi-choice component consisting of tracing code, explaining code, testing code, and debugging, along with short answer and Parsons Problems, 2) (45 minutes, worth 15\% of the grade) four code writing tasks of increasing difficulty to be completed without Copilot, 3) (45 minutes, worth 15\% of the grade) one large new problem to be completed with Copilot.  For the first two parts of the exam, students worked on a lab machine with access to a web browser for PrairieLearn.  For the third part of the exam, students could use Copilot either on their personal computer or in a PrairieLearn workspace we provided.  Proctors ensured students were only using allowed content.

For the third part of the exam, students were asked to analyze a new dataset or implement a restricted version of a spell check. 
For example, for spell check, students needed to take a given word and return all correctly spelled words that can be reached by adding, removing, or changing one character.  Students were given example test cases and were provided partial credit for partial progress.

\subsectionhack
\subsectionhack
\section{Student Perceptions}

Given the scope of changes we made to CS1, we wanted to understand the impacts on the student experience. To that end, we asked students for their opinions about the course in an end-of-course survey.  We report here on questions from that survey that directly relate to the student perceptions of working with an LLM.

For the quantitative questions, we created graphs to visualize the responses. For the open-ended questions, we identified quotes in the data that provided context to these initial visualizations and findings. We did not conduct a formal quantitative or qualitative analysis of the data at this point.

\subsection{Student Comfort with Copilot}
\subsectionhack

\begin{figure}[ht]
    \centering
        \includegraphics[width=\columnwidth]{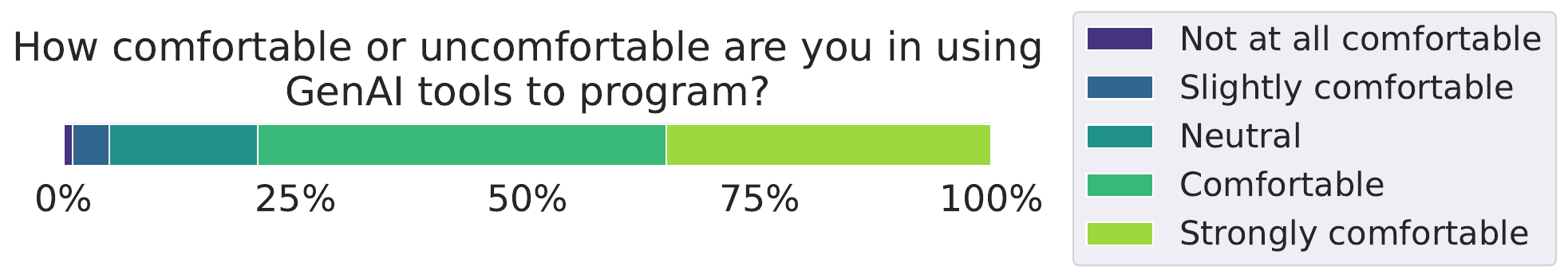}
    \vspace{-5mm}
    \caption{Student Comfort Programming with GenAI\figendhack}
    \label{fig:comfort}
\end{figure}

We asked students: ``How comfortable or uncomfortable are you in using GenAI tools to program?'' and their responses appear in Figure~\ref{fig:comfort}.  Encouragingly, the vast majority (79\%) reported being comfortable using GenAI tools to program.

\subsection{Impact of Copilot on Student Learning}
\subsectionhack

\begin{figure}[ht]
    \centering
        \includegraphics[width=\columnwidth]{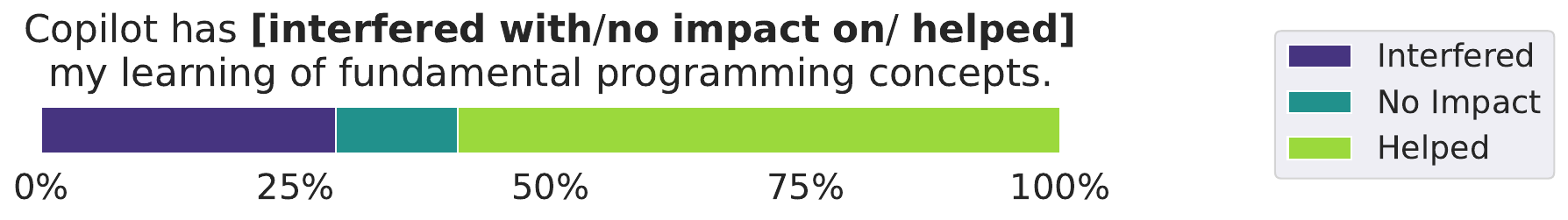}
    \vspace{-5mm}
    \caption{Student Perceptions of how Copilot impacted their learning.\figendhack}
    \label{fig:confidence_summary}
\end{figure}

When asked about their overall opinion of working with Copilot, a slight majority (59\%) of students report that it helped their learning of programming concepts (Figure~\ref{fig:confidence_summary}). 
 Student responses to an open-ended survey question (``How did you feel about working with Copilot as you learned to program this quarter?'') offered insights
 into why students reported that Copilot either helped or interfered with their learning.  Among students who reported that it was helpful for their learning, some
 students reported an appreciation for having immediate personalized help.  For example, one student reported that ``It was really nice having an assistant that could always help me the moment I needed it and made programming a lot less daunting.''  Among students who reported
 that it interfered with their learning, many reported that they found Copilot useful but felt they had become over-reliant on it. For example, ``Copilot allowed me to develop a sufficient understanding of a lot of concepts, but I wasn't necessarily able to master most of those concepts. I feel like this is because Copilot enhances the speed I'm able to learn at, but doesn't encourage me to master the concept to the level where I'm able to write the code entirely on my own...'' Another student reported ``If I were asked to code without Copilot, I wouldn't feel very confident in myself despite doing well in the course.'' 

\begin{figure}[ht]
    \centering
    \includegraphics[width=\columnwidth]{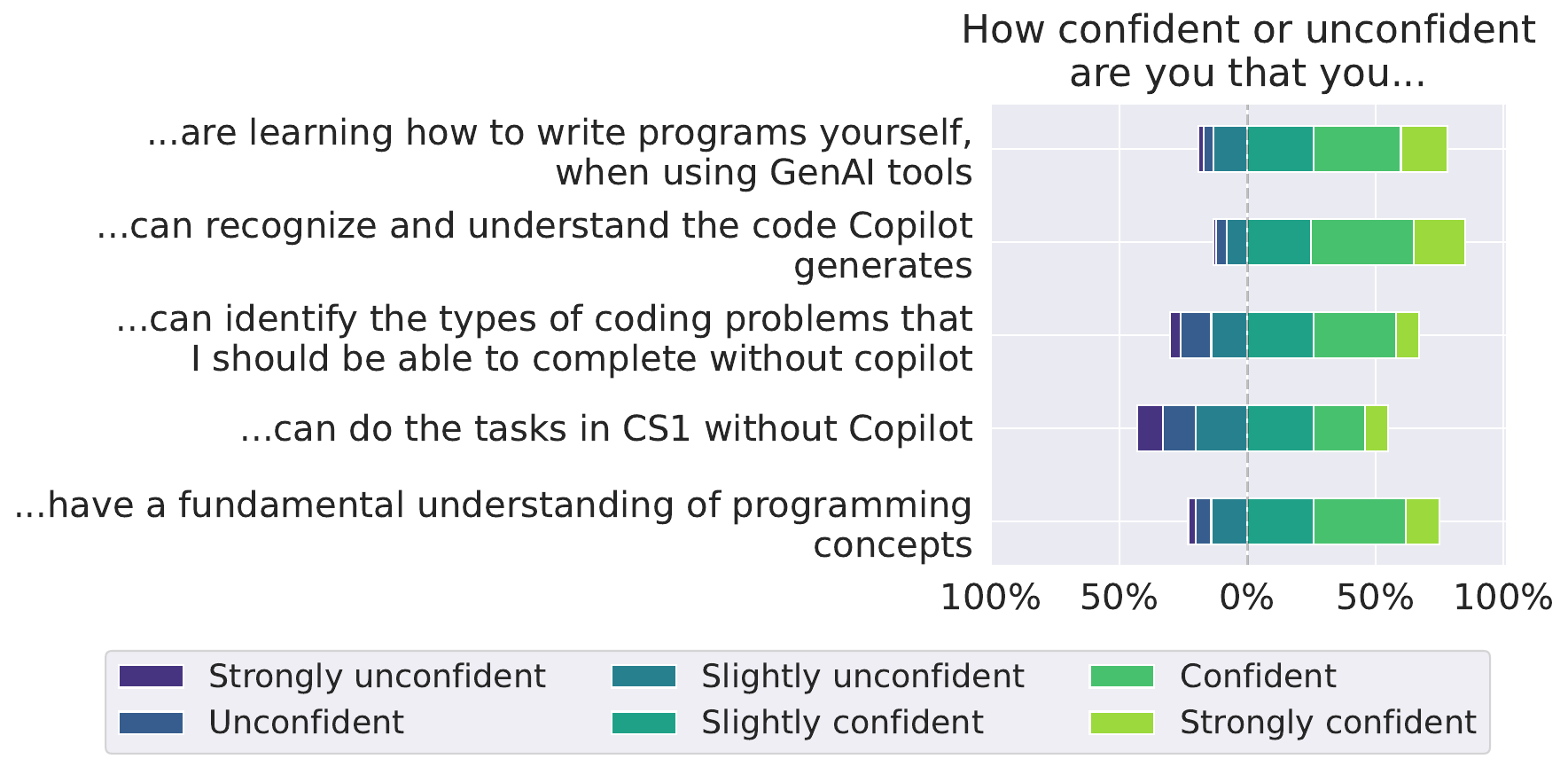}
    \vspace{-6mm}
    \caption{Student confidence in their ability and understanding at the end of the course.\figendhack}
    \label{fig:confidence_detailed}
\end{figure}

We gain a deeper insight into this range of opinions through the questions asked in Figure~\ref{fig:confidence_detailed}.  Students overall felt confident that they are learning how to write programs on their own when using GenAI tools, that they could recognize and understand the code generated by Copilot, and that they have gained a fundamental understanding of programming concepts.  Fewer students reported confidence in their ability to perform the tasks from the course without Copilot.  This may be related to their lack of confidence
 in their ability to ``identify the types of coding problems they should be able to complete without Copilot''; 31.1\% of the students were slightly to strongly unconfident that they could do so (Figure ~\ref{fig:confidence_detailed}). Some students felt frustrated by their inability to use Copilot for some portions of the course and not others.  A representative quote from a student is that ``Although it was helpful, it was really difficult for me on quizzes, the whole course is based on using AI tools such as copilot for help, yet taking it away on quizzes when we had it in every other type of work seemed a little unfair.''  As we discuss in Section~\ref{sec:lessons_learned}, we could have offered more guidance about what we expect students should be able to code with and without Copilot.
 
\subsectionhack
\subsection{How Students Interacted with Copilot}

\begin{figure}[ht]
    \centering
    \vspace{-2mm}
    \includegraphics[width=\columnwidth]{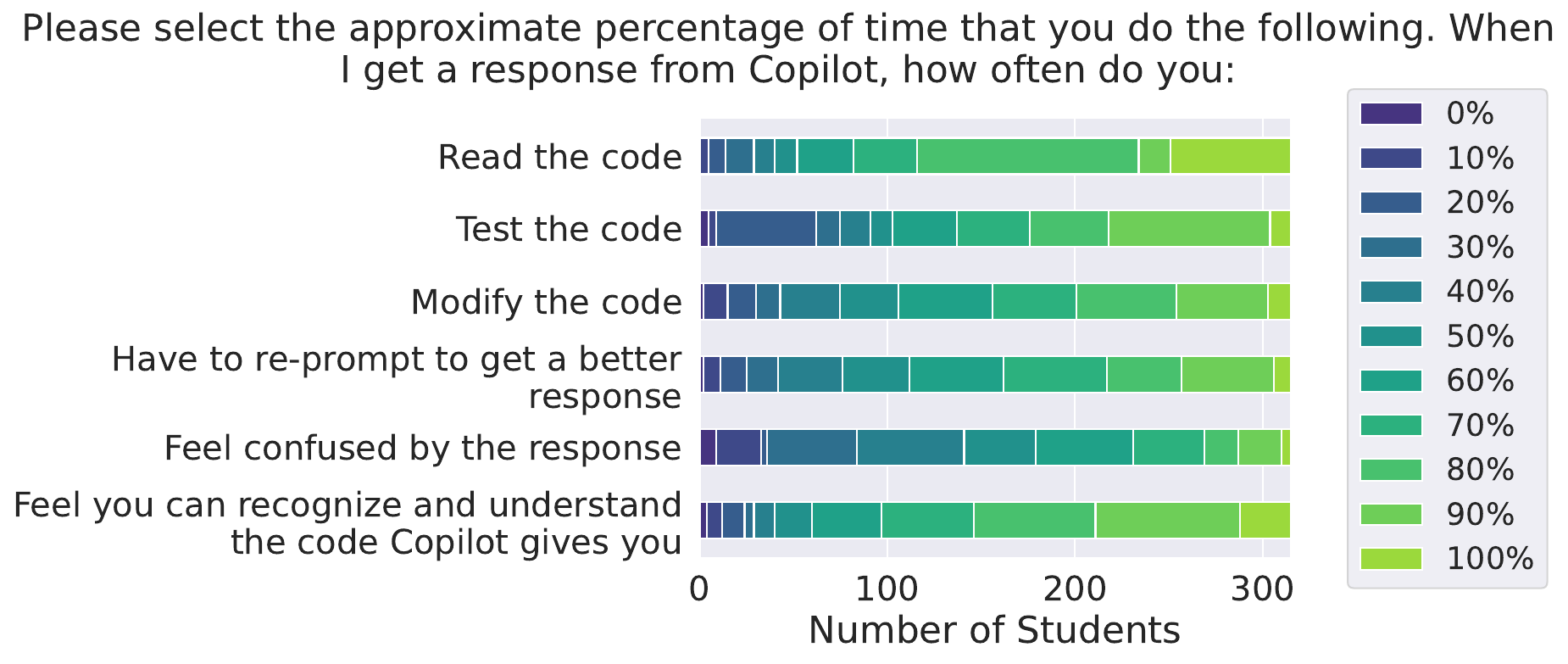}
    \vspace{-5mm}
    \caption{Student perceptions on how they interact with Copilot.\figendhack}
    \label{fig:percentage_plot}
\end{figure}

Figure~\ref{fig:percentage_plot} summarizes how students worked with Copilot during the term.  We're encouraged that the majority of students said that they read the code returned by Copilot at least 80\% of the time, and that they tested and modified the code at least 60\% of the time.  
It is not surprising that students occasionally or even frequently did not have to modify the code returned by Copilot; indeed, we coached them on how to give prompts that would lead to code that was as close to what they were looking for as possible.  
Finally, while around a third of students reported only rarely (30\% of the time or less) being confused by Copilot's responses, 44\% of students reported being confused 50\% of the time or more, with 10\% of students
reporting being confused the vast majority of the time (80\% of the time or more).  Perhaps a reason for this trend is that Copilot can sometimes give sophisticated solutions that are beyond the scope of an introductory course, but we have not yet analyzed the complexity of Copilot responses.

\subsectionhack
\subsectionhack
\subsection{Student Perceptions of the Projects}

\begin{figure}[ht]
    \vspace{-2mm}
    \centering
        \includegraphics[width=\columnwidth]{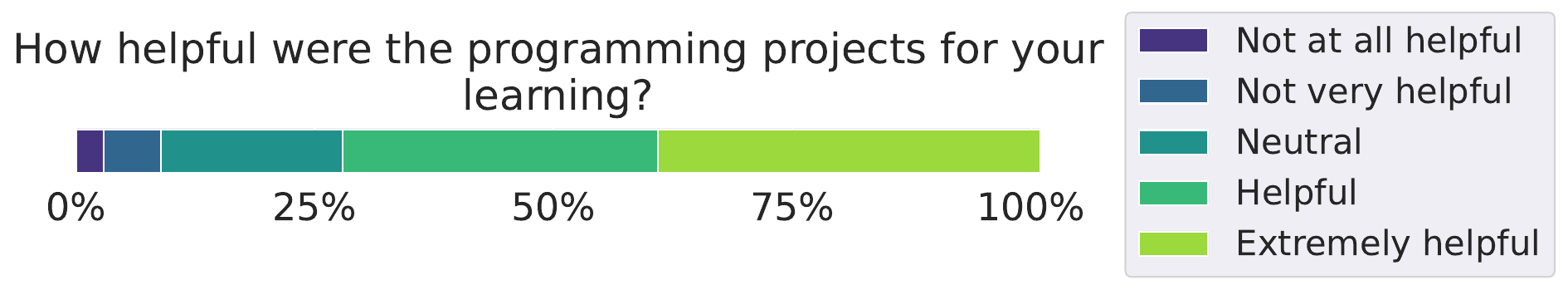}
    \vspace{-6mm}
    \caption{Students' View on Learning with Projects\figendhack}
    \label{fig:project_learning}
\end{figure}

We also asked students about the projects, as they played a large role in the new class.  These projects were designed to be open-ended and more challenging than typical CS1 projects so students could showcase their ingenuity. 
Students largely reported that these projects were valuable for their learning, as shown in Figure~\ref{fig:project_learning}. 

Students overall seemed to enjoy projects, reflected in one representative comment: ``Though they were a little frustrating at times, I was really impressed by what I was able to do for each project and they are probably what I will remember most from the class.  I [...] enjoyed being able to be creative and apply what we learned in class. I also thought they were challenging enough, while still being doable.  I definitely felt more comfortable with [...] coding in general after each project.''

\subsectionhack
\section{Discussion}
\label{sec:lessons_learned}

As shown in the last section, a vast majority (79\%) of students felt they could program with the LLM and a slight majority (59\%) felt it was helpful in learning programming.  In contrast, some students reported concerns about an over-reliance on the tool or felt they may not have learned the fundamentals as well as they'd hoped. While these concerns should be addressed, students did create projects far beyond what we would have previously expected in a CS1 course. Thus, it is not surprising that some students felt like they are not capable of writing the code without an LLM, as a CS1 student would not be expected to write such advanced code. 

The changes we made to the course led to both design and administrative challenges. That said, we are encouraged by the student enthusiasm for the new course and open-ended projects. We plan to iterate on the course and are optimistic about future offerings.

For instructors looking to incorporate LLMs into their teaching of CS1, we offer these observations and lessons learned based on the experiences of the instructional staff and student feedback.

\noindent \textbf{Student Performance.}  
Student performance on exams mostly mirrored performance in previous CS1 classes.  Anecdotally, the CS1-LLM students' performance on code writing (from scratch) questions was slightly lower than past offerings, but their performance on code tracing and code reading questions was roughly the same. As already discussed, the scope of their projects was well beyond what we would regularly see in a CS1 course. We were also pleased to see that many students were able to solve the large programming task in the 3rd part of the final exam. Future studies will explore student performance in more detail.

\noindent \textbf{Essential Components.}  
If one is hesitant to fully adopt a new course, the changes we feel are most essential include the projects (on which students used LLMs), teaching of problem decomposition and testing, and using LLM code responses in class examples.

\noindent \textbf{Extent of Changes.}  Although nearly all elements of the course were changed, ultimately the depth of the changes felt less than we initially expected. Students still learned how to read, trace, explain, and write basic code.  The larger changes occurred when interacting with the LLM in lecture, teaching problem decomposition and testing, using larger projects, and assessing students with Copilot.

\noindent \textbf{When to Introduce LLMs.}  We introduced LLMs in the first week. This meant that students were grappling with using an IDE, using an LLM, and working with Python all at the same time. In future offerings, we recommend delaying introducing the LLM briefly so students can write small programs in the IDE on their own first.

\noindent \textbf{Explicit Expectations.}  A common concern voiced in surveys was that students were confused about what they should be able to do with and without the aid of an LLM.  In hindsight, the advice that students could use LLMs if stuck on homework may not have been helpful as some students reported using Copilot heavily for the homework and becoming over-reliant.  In future offerings, we will tag every homework question clearly as something they should be able to do with/without the aid of an LLM.

\noindent \textbf{Learning Goals and Assessments.} A challenge in re-imagining a course is in ensuring the assessments are aligned to the learning goals of the course.  Given the many decades of teaching CS1 without the aid of an LLM, our team often defaulted back to the kinds of questions we asked in prior versions of the course. 
If past questions are used, we recommend checking that the questions align with the (updated) course learning goals.

\noindent \textbf{Beware of Non-Determinism in Class.}  A challenge in live-coding with an LLM is that it learns from your behavior.  In teaching two sections of the same class, the LLM made discussion-worthy mistakes in the first section but then simply parroted back our fixed code from that section for the second section. We learned to have Copilot responses pasted into our slides rather than Live Coding.

\noindent \textbf{Signing into GitHub is Difficult in Exams.}  
For part of Quiz 2, we attempted to have students solve a small task in 15 minutes with the aid of Copilot in a PrairieLearn workspace with Copilot enabled.  Students were unable to finish this task as much time was lost due to forgotten credentials or need of their phones for 2-factor verification.  In general, if students are allowed Copilot during an exam, we recommend allotting sufficient time to log in or allowing them to use a personal device with stored credentials.

\subsectionhack
\section{Conclusion}

We offered a new kind of introductory programming course (CS1-LLM) that integrated LLMs into the course instruction.  We described the revised learning goals, the course structure, and lessons learned for potential course adopters.  From student surveys and student projects, we learned how students interact with the LLM, that students in the course valued the open-ended course projects, and that the majority felt that the LLM helped their learning. 
We see this course offering as a first step toward using LLMs to improve student outcomes and experience in CS1.

\subsectionhack
\section{Acknowledgements}

We appreciate the reviewers for their helpful feedback as well as funding from Google's Award for Inclusion Research.  
\bibliographystyle{ACM-Reference-Format}
\bibliography{references}

\end{document}